\documentclass[%
 aip,
 amsmath,amssymb,
 reprint,%
]{revtex4-1}
\usepackage{float}
\usepackage{graphicx}
\usepackage{dcolumn}
\usepackage{bm}
\usepackage{xcolor}
\usepackage[utf8]{inputenc}
\usepackage[T1]{fontenc}
\usepackage{mathptmx}
\usepackage{etoolbox}
\usepackage{booktabs}
\usepackage{xr}

\makeatletter
\def\@email#1#2{%
 \endgroup
 \patchcmd{\titleblock@produce}
  {\frontmatter@RRAPformat}
  {\frontmatter@RRAPformat{\produce@RRAP{*#1\href{mailto:#2}{#2}}}\frontmatter@RRAPformat}
  {}{}
}%
\makeatother

\usepackage{xr-hyper}
\usepackage{hyperref}
\begin{document}

\preprint{AIP/123-QED}

\title{Benchmarking distinguishable cluster methods to platinum standard CCSDT(Q) non covalent interaction energies in the A24 dataset}

\author{S. Lambie}
\affiliation{Max Planck Institute for Solid State Research, Heisenbergstr. 1, 70569 Stuttgart, Germany}
 
 \author{C. Rickert}
 \affiliation{Institut f\"ur Chemie, Humboldt-Universit\"at zu Berlin, Brook-Taylor-Str. 2, Berlin 12489, Germany}
 
 \author{D. Usvyat}
 \affiliation{Institut f\"ur Chemie, Humboldt-Universit\"at zu Berlin, Brook-Taylor-Str. 2, Berlin 12489, Germany}

\author{A. Alavi}
\affiliation{Max Planck Institute for Solid State Research, Heisenbergstr. 1, 70569 Stuttgart, Germany}%
\affiliation{Yusuf Hamied Department of Chemistry, University of Cambridge, Lensfield Road, Cambridge CB2 1EW, United Kingdom}%

\author{D. Kats}%
\affiliation{Max Planck Institute for Solid State Research, Heisenbergstr. 1, 70569 Stuttgart, Germany}%

\email{s.lambie@fkf.mpg.de; d.kats@fkf.mpg.de}

\date{\today}

\begin{abstract}
Recent disagreement between state-of-the-art quantum chemical methods, coupled cluster with single, double and perturbative triples excitations and fixed-node diffusion Monte Carlo, calls for systematic examination of possible sources of error within both methodological approaches. Coupled cluster theory is systematically improvable toward the exact solution of the Schr\"odinger equation, however very quickly is limited by the computational cost of the calculation. Therefore, it has become imperative to develop low-cost methods that are able to reproduce CC results, beyond the CCSD(T) level of theory. Here, the DC-CCSDT and SVD-DC-CCSDT methods are examined for their fidelity to the CCSDT(Q) correlation interaction energies for the A24 dataset and are shown to outperform CCSDT and CCSD(T). Furthermore, with (T)-based corrections of the SVD approximation the SVD-DC-CCSDT method becomes an accurate and relatively low-cost tool for calculation of previously intractable post-CCSD(T) energies in atomic orbital basis sets of unprecedented size.  
\end{abstract}

\maketitle

\begin{quotation}

\end{quotation}

\section{Introduction}

Coupled cluster (CC) theory with single, double and perturbative triples (CCSD(T)) is the gold standard quantum chemical approach for the vast majority of chemical applications.\cite{helgaker:2000} However, the reports of disagreement between fixed-node diffusion Monte Carlo results and CCSD(T) for non covalent interaction energies\cite{al-hamdani:2021, al-hamdani:2019, ballesteros:2021} necessitate that higher order excitations are included in the cluster operator of CC to more systematically benchmark the performance of CCSD(T) for inter-molecular interaction energies. It has previously been shown that one should go to at least the CCSDT(Q) level of theory to obtain converged results with respect to the cluster operator.\cite{rezac:2013, rezac:2013b, demovicova:2014} Unfortunately, the already steep scaling of conventional CCSD(T) ($\mathcal{O}$(\textit{N}$^7$)) and the necessity of including higher order excitations in the cluster operator make obtaining highly accurate interaction energies a difficult endeavor. As such, there is a real need to develop highly accurate, systematically improvable methods to approximate CC methods beyond the CCSD(T) level of theory. 

There exist many different lines of inquiry for developing low cost, high accuracy methods to approximate CCSD(T) or better results. Approaches to reduce the computational cost of quantum chemical calculations extend from linear-scaling local CC methods: pair natural orbital local CC,\cite{ma:2018} local pair natural orbital CC\cite{neese:2009} and domain local pair natural orbital CC,\cite{nagy:2018} to tensor decomposition approaches such as density fitting,\cite{feyereisen:1993, rendell:1994} density matrix renormalization group\cite{schollwock:2005} and singular value decomposition\cite{hino:2004, kinoshita:2003, lesiuk:2020} techniques. Furthermore, the conventional CC results can be approximated through alternative CC-based theories such as distinguishable cluster,\cite{kats:2013, kats:2014, kats:2019} CCSD(cT)\cite{masios:2023, schafer:2024} or other methods.\cite{paldus_approximate_1984,piecuch_solution_1991,
bartlett_addition_2006,nooijen_orbital_2006, Neese:09, huntington_pccsd:_2010,robinson_approximate_2011,
huntington_accurate_2012,paldus_externally_2016,black_statistical_2018,teke_best_2024}

Here, we focus on distinguishable cluster\cite{kats:2013,kats:2014} (DC) CCSDT\cite{kats:2019,rishi_can_2019,schraivogel:2021} and singular value decomposed (SVD) DC-CCSDT.\cite{rickert:2025} DC-CCSDT reduces the cost of CCSDT by selectively removing some exchange terms.\cite{kats:2019} For reaction energies, thermochemical properties and excited states for both open and closed shell species, DC-CCSDT has been shown to be more accurate than conventional CCSD(T) and CCSDT when compared to CCSDT(Q).\cite{schraivogel:2021, kats:2019} 
The SVD-DC-CCSDT approach was introduced to further reduce the cost and memory of the DC-CCSDT method using tensor decomposition to reduce the size of the CC amplitudes while, simultaneously, retaining the excellent performance of the DC-CCSDT method. Benchmark calculations for a test set of 68 reaction energies found that the error associated with the SVD approximation was small when sufficiently tight SVD amplitude thresholds were used
or when (T)-based corrections were applied (either using full (T) energies, denoted SVD-DC-CCSDT+, or using SVD-(T) energies with tighter thresholds, denoted SVD-DC-CCSDT*).\cite{rickert:2025}

While the performance of the DC-family of quantum chemical methods is promising, testing is not yet complete. Interaction energies of molecular systems is, perhaps, one of the most rigorous tests of a method's performance because these interactions are usually minute at the scale of the total energies and thus are sensitive to energetic changes up to the micro-Hartree regime. 
To quantify the performance of novel methods, high accuracy benchmark datasets are required -- of which many exist, including  the A24,\cite{rezac:2013b} X40,\cite{rezac:2012} S22,\cite{jurecka:2006} S66,\cite{rezac:2011} L7\cite{sedlak:2013} and S12\cite{grimme:2012} datasets. However, only two datasets currently report non covalent interaction energies beyond the CCSD(T) level of theory; the A24 dataset\cite{rezac:2013b, rezac:2015} and the most recent iteration of benchmark calculations on the S66 dataset.\cite{semidalas:2025} For the A24 dataset, the interaction energies were calculated at the CCSDT(Q)/aug-cc-pVDZ level of theory while the benchmark S66 values were calculated using CCSDT(Q)/cc-pVDZ. Augmentation functions are known to be vital for accurately capturing non-covalent interactions\cite{sinnokrot:2004, marshall:2011, witte:2016, nagy:2019} and, as such, we select the A24 dataset CCSDT(Q)/aug-cc-pVDZ results as platinum standard interaction energies (albeit, still in a small basis set) to which the DC-CCSDT and SVD-DC-CCSDT methods are benchmarked for non covalent interactions.

The A24 dataset consists of 24 non covalent dimers encompassing H-bonding (systems 1-5), mixed electrostatic and dispersion (systems 6-15) and predominantly dispersion (systems 16 - 24) interactions (Figure \ref{structures}). Geometries range in size from a maximum of 24 correlated electrons (ethene dimer, system 23), to a minimum of 14 correlated electrons (borane-methane dimer, system 16), within the frozen core approximation. Systems 1-21, inclusive, are optimized at a composite CCSD(T)/complete basis set limit level of theory\cite{rezac:2013b} and sit at a minimum on the potential energy surface. Systems 22-24, however, were included to be representative of $\pi-\pi$ stacking interactions and these geometries are not minima on the potential energy surface, but were instead fixed at a distance of 3.5 \AA{} in a fully symmetric (`sandwich') arrangement. 

In this communication, 
we show that SVD-DC-CCSDT+ and SVD-DC-CCSDT* with tight SVD amplitude cutoffs perform excellently compared to platinum standard CCSDT(Q) calculations. It is also demonstrated that these methods require a fraction of the cost of conventional DC-CCSDT and enable calculation of interaction energies in atomic orbital basis sets that have so far been unfeasible for such a level of theory.

\section{Singular value decomposed Distinguishable Cluster theory}\label{sec:theory}

In this section we provide a brief overview of the SVD-DC-CCSDT method; the detailed theory can be found in Refs. \citenum{rickert:2025}. The distinguishable cluster approach is generally based on coupled cluster (CC) theory,\cite{civzek:1966, bartlett:2007, bartlett:1991, bartlett:2024, crawford:2000, bishop:1991} in which the wavefunction is parameterized via an exponential ansatz
\begin{eqnarray}
      \left| \Psi_{CC} \right> = e^{\hat{T}} \left| \psi_{0} \right>,\label{eq:exp}
\end{eqnarray}
with the cluster operator
\begin{eqnarray}
    \hat{T} = \hat{T}_1 + \hat{T}_2 + \hat{T}_3 + \ldots,\label{eq:T}
\end{eqnarray}
defined as
\begin{eqnarray}
    \hat{T}_n = \left( \frac{1}{n!} \right)^2 T_{AB\ldots}^{IJ\ldots}   \hat{a}_A^\dag \hat{a}_I\hat{a}_B^\dag \hat{a}_J\ldots . \label{eq:clust}
\end{eqnarray}
Here, $\hat{a}^\dag$ and $\hat{a}$ are the 
creation and annihilation operators, respectively, with the indices $I,J,K, ...$ denoting the occupied spin-orbitals and  $A,B,C, ...$ 
-- the virtual ones, $T_{AB\ldots}^{IJ\ldots}$ are the corresponding excitation amplitudes. The exponential operator in eq. (\ref{eq:T}) is applied to a reference function $\psi_{0}$, commonly chosen as the Hartree Fock Slater determinant.
In eq. (\ref{eq:clust}) and in the following, 
summations over repeated indices are assumed.
The projected linked CC energy and amplitude equations read
\begin{eqnarray}
  \left< \psi_{0} \left| e^{- \hat{T}} \hat{H}_N e^{\hat{T}} \right| \psi_{0} \right> &=&E_{corr}\label{eq:en}\\
 \left< \psi^{I\ldots}_{A\ldots} \left| e^{- \hat{T}} \hat{H}_N e^{\hat{T}} \right| \psi_{0} \right> &=& 0,\label{eq:ampl} 
   \end{eqnarray}
where  $\hat{H}_N$ is the normal ordered Hamiltonian and $E_{corr}$ is the correlation energy.
The crucial feature of the exponential form of the wavefunction (eq. \ref{eq:exp}) is its multiplicative factorization, which even after truncation of the cluster operator at a certain excitation level, guarantees the size-extensivity of the correlation energy.

The DC approximation can be seen as an alternative way  to truncate the full CC expansion compared to the simple truncation according to the excitation level, as in the standard CC theory.\cite{kats:2013, kats:2014} The DC approaches omit certain quadratic exchange terms in the CC amplitude equations and rescale other quadratic terms to maintain the size-extensivity, orbital invariance, particle-hole symmetry\cite{kats:2018} and exactness for an $n$-electron system, where $n$ is the truncation level. 
The exact definition and the formalisms of
DCD, DCSD, Brueckner DCD, orbital-optimized DCD and DC-CCSDT can be found in Refs. \citenum{kats:2013, kats:2014, schraivogel:2021, rishi:2019}, respectively.  
Interestingly, despite the modified and shortened amplitude equations, DC is known to uniformly outperform standard CC within a given truncation level.\cite{kats:2015,kats:2018a,khomyakovReliable2017,kesharwaniSurprising2017,tsatsoulis_comparison_2017,blackStatistical2018,petrenkoRefined2018,limanniRole2019,kats:2019,woller_performance_2020,lin_fragment-based_2020,schraivogel:2021,vitale_fciqmc-tailored_2022,schraivogel_two_2024} 
Particularly for DC-CCSDT, which is relevant for this work,  it was recently shown on model systems that compared to CCSDT it is closer to the CCSDT(Q) reference for reaction energies.\cite{kats:2019, schraivogel:2021}

Although DC-CCSDT is computationally less expensive than CCSDT (e.g. in the real-space representation its nominal scaling is $\mathcal{O}(N^7)$\cite{kats:2019} vs  $\mathcal{O}(N^8)$ of CCSDT) without 
further approximations it is hardly applicable even to medium sized systems. 
To remedy this, a SVD approximation has been recently developed for DC-CCSDT, which allowed for a substantial boost in efficiency.\cite{rickert:2025}

In the SVD-DC-CCSDT approach, the triples amplitudes $T^{ijk}_{abc}$ (the indices $i,j,k, \dots$ and $a,b,c, \dots$ denote the occupied and virtual spacial orbitals, respectively) are decomposed as 
\begin{eqnarray}
    T^{ijk}_{abc} &=& T_{XYZ} \: U^{iX}_a U^{jY}_b U^{kZ}_c.\label{eq:trip_decomp}
  \end{eqnarray}
The symmetric three-index quantity $T_{XYZ}$ can be interpreted as the SVD-basis representation of the triples 
amplitudes.  The transformation matrices  $U^{iX}_a$ between the SVD space and the space of occupied-virtual orbital pairs are the eigenvectors of an approximate triples two-particle density matrix $D_{ia,bj}$. In our approach  we use a CC3-type density matrix for the latter, which is calculated from the CCSD doubles amplitudes for a $\mathcal{O}(N^6)$ cost; see Ref. \citenum{rickert:2025} for details.

Using the decomposition (eq. \ref{eq:trip_decomp}) one can formulate  the triples amplitude equations such that the triples amplitudes or residuals, which are six-index tensors in the orbital basis, remain entirely in the SVD space. However, the exact decomposition per se does not bring a cost reduction as the size of the SVD basis is equal to the number of occupied-virtual orbital pairs. To exploit the actual effectiveness of this approach the SVD basis should be truncated. This can be done, for example,  by neglecting the eigenvectors of $D_{ia,bj}$, corresponding to small enough eigenvalues. This cuts down the size of the SVD basis dramatically, leading to a very compact set of the amplitudes and residuals and bringing large savings in memory and storage. Furthermore, as is demonstrated in Ref. \citenum{rickert:2025} on a model alkane chain, the size of the truncated SVD basis using a cutoff threshold for the eigenvalues of $D_{ia,bj}$ grows linearly with the system size. With the density fitting factorization of the two-electron integrals, this lowers the overall scaling of the triples residual equations in SVD-DC-CCSDT to $\mathcal{O}(N^6)$, as well as significantly reducing the prefactor.

In our approach, the cutoff threshold that defines the SVD-approximation in the SVD-DC-CCSDT method, is compared to the eigenvalues of the approximate two-electron density matrix  $D_{ia,bj}$.\cite{rickert:2025} By tightening this threshold one can progressively release the SVD approximation and converge to the true DC-CCSDT results. In the following, we systematically investigate how the choice of this threshold, which we refer to as the `SVD amplitude threshold' throughout this paper, affects the accuracy and efficiency of the SVD-DC-CCSDT method for intermolecular interaction energies.

It can be expected that the error of the SVD-truncation can be further reduced if a part of the triples contribution to the correlation energy is calculated without this approximation. As is demonstrated  Ref. \citenum{rickert:2025} a simple (T)-correction scheme, where the difference between (T) and SVD-(T) energies is added to the SVD-DC-CCSDT result, indeed improves the accuracy by one or more orders of magnitude in the SVD-threshold. The computational cost of (T) even without the SVD approximation is rather moderate compared to SVD-DC-CCSDT, at least for tighter SVD amplitude thresholds, making this scheme quite efficient. We denote the method  which employs this correction as ``SVD-DC-CCSDT+''. 

The scaling of non-SVD-(T) is $\mathcal{O}(N^7)$, which is higher than that of SVD-DC-CCSDT. This means that evaluation of (T) may asymptotically become the bottleneck in the SVD-DC-CCSDT+ scheme. In order to eliminate the need for a $\mathcal{O}(N^7)$-step
altogether we also test a correction scheme involving SVD-(T)
energies only, whereby the correction is obtained from the
difference between the SVD-(T) energy using the threshold of the SVD-DC-CCSDT
calculation and the SVD-(T) energy obtained by tightening the aforementioned threshold by one or two orders of magnitude. This approach
is denoted as ``SVD-DC-CCSDT*''.


\section{Computational Details}

\begin{figure*}
\centering
\includegraphics[width=1.0\textwidth]{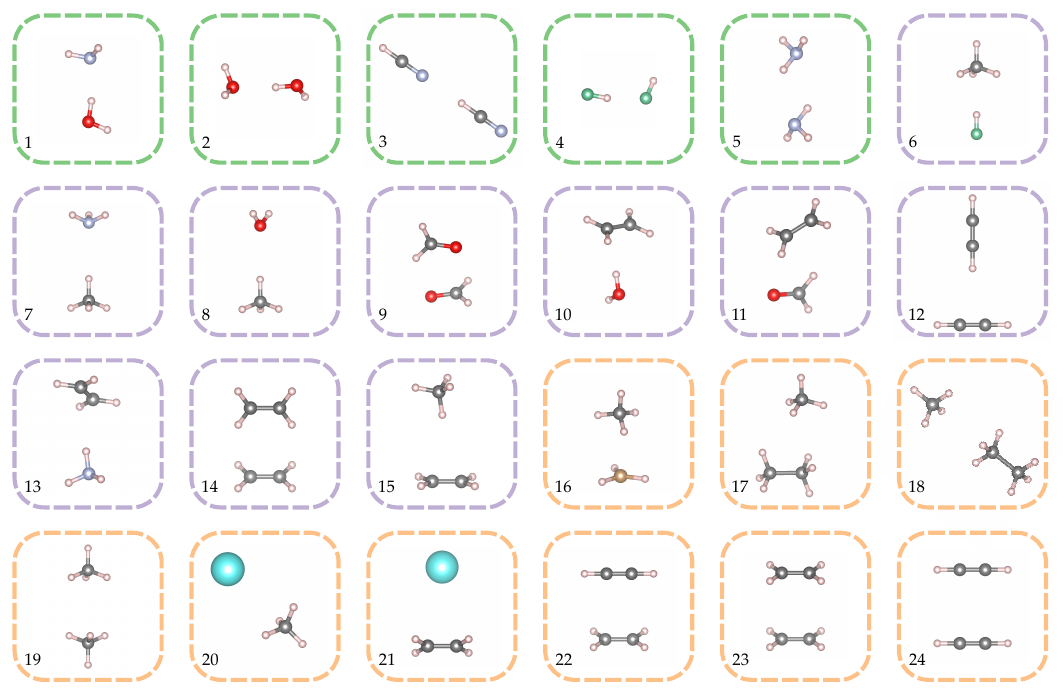}
\caption{Systems included in the A24 dataset. Systems 1-5 interact through hydrogen bonding (green boxes), systems 6-15 interact through mixed electrostatic and dispersion interactions (purple boxes) and systems 16-24 interact predominantly through dispersion (orange boxes). Atom types are colored according to the Corey–Pauling–Koltun coloring conventions.}
\label{structures}
\end{figure*}

Geometries are taken from the Benchmark Energy and Geometry Database (Figure \ref{structures}).\cite{begd, rezac:2008} All calculations are carried out in \texttt{ElemCo.jl}\cite{elemcoil} using the counterpoise correction scheme of Boys and Bernardi\cite{boys:1970} to help with addressing the basis set superposition error. The frozen core approximation is also employed throughout this study.

\subsection{Density fitting testing}

Testing is carried out in Section \ref{df-testing} to determine the required size of the JKFIT\cite{weigend:2002} and MPFIT\cite{weigend:2002b} density fitting basis sets as to minimize the error introduced by density fitting while not increasing the computational cost unnecessarily. The JKFIT basis sets are varied using the correlation consistent cc-pV\textit{X}Z basis sets\cite{weigend:2002} where \textit{X} = D, T, Q or 5 for all atoms within a system, with the exception of Ar, for which the correlation consistent JKFIT basis sets are unavailable. The def2-universal JKFIT\cite{weigend:2008} basis set is used for the Ar atoms in systems 20 and 21. The atomic orbital basis set used for all density fitting test calculations is aug-cc-pVDZ.\cite{dunning:1989} For benchmarking the density fitted results, the reference SCF/aug-cc-pVDZ and CCSD(T)/aug-cc-pVDZ counterpoise corrected correlation interaction energies are taken from Burns \textit{et al.}\cite{burns:2014} Ultimately, once the testing was completed, a JKFIT of cc-pVQZ and MPFIT of aug-cc-pVTZ for the density fitting basis set sizes is used for all remaining density fitted calculations carried out. 

\subsection{Reference values}

The CCSDT(Q)/aug-cc-pVDZ reference values are taken from \v{R}ez\'{a}\v{c} \textit{et al.}\cite{rezac:2015} and the CCSD(T)/aug-cc-pVDZ values are taken from Burns \textit{et al.}\cite{burns:2014} These reference values are used to benchmark the DC-CCSDT and SVD-DC-CCSDT methodologies. We highlight that system 18 is excluded from benchmarking to CCSDT(Q) reference values as this system was not reported in reference \citenum{rezac:2015} due to computational limitations. Additionally, density fitted CCSDT/aug-cc-pVDZ correlation interaction energies are calculated for all systems, excepting 18, to compare the performance of CCSDT to the CCSDT(Q) reference values. The density fitted CCSDT calculations use JKFIT\cite{weigend:2002} of cc-pVQZ and MPFIT\cite{weigend:2002b} of aug-cc-pVTZ. 

\subsection{Distinguishable cluster calculations}


Two additional variants of the SVD-DC-CCSDT method are available: SVD-DC-CCSDT* and SVD-DC-CCSDT+. The `+' and `*' variants reduce the SVD approximation error by adding an SVD-(T)-error correction, see section \ref{sec:theory} and Ref. \citenum{rickert:2025} for details.  To systematically approach the DC-CCSDT result, the values of the SVD threshold are tested in the range from $10^{-5}$ to $10^{-7}$.

\section{Results and Discussion}

\subsection{Benchmarking density fitting}\label{df-testing}

We begin by benchmarking the JKFIT and MPFIT auxiliary basis sets used within the density fitted implementations of the quantum chemical approaches to their conventional counterparts. The JKFIT is used in the construction of the Fock matrix\cite{weigend:2002} while the MPFIT (also called RIFIT or MP2FIT in other codes) is used for all other integrals within the calculation.\cite{weigend:2002b} 
As DC-CCSDT, SVD-DC-CCSDT, SVD-DC-CCSDT*, SVD-DC-CCSDT+ and CCSDT calculations, reported in sections \ref{dc-ccsdt}-\ref{comp-time}, all employ density fitting, it is important to assess and minimize the density fitting error such that it becomes negligible compared to the method error. 

\subsubsection{Hartree-Fock interaction energy}

\begin{figure}
		\centering
		\includegraphics[width = 0.45\textwidth]{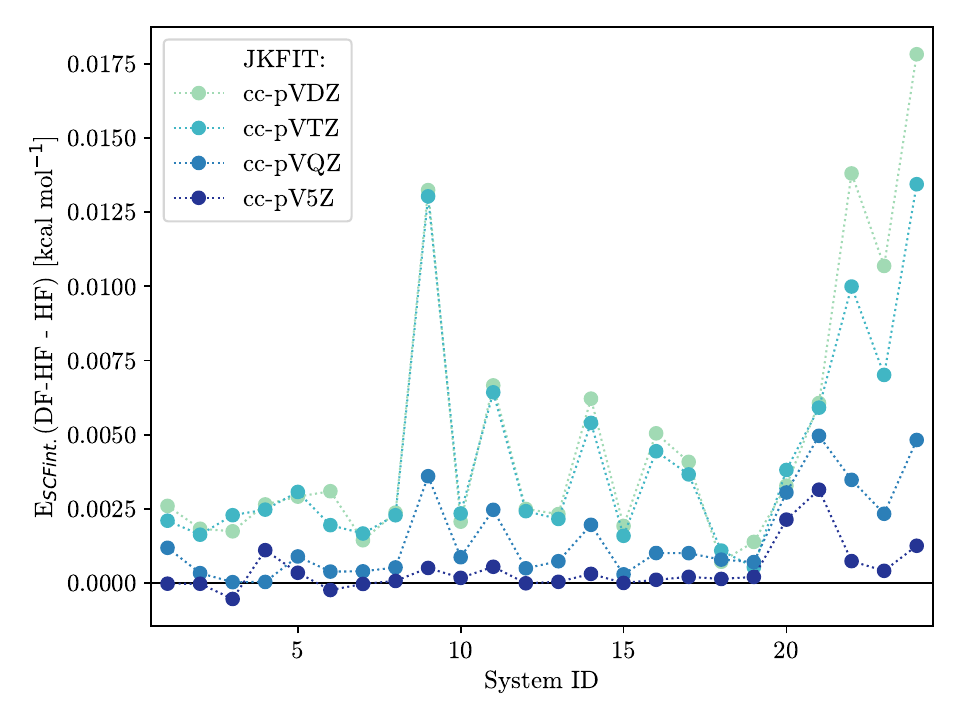}\\
		\caption{Error resulting from density fitting compared to the non-density-fitted result, for the SCF interaction energy as a result of changing the JKFIT density fitting basis set size. }
		\label{df-hf}
	\end{figure}

 In the first instance, we compare the non-density-fitted SCF interaction energy between two monomers to that obtained with JKFIT density fitting (Figure \ref{df-hf}). Figure \ref{df-hf} and Table \ref{df-hf-err} demonstrate that there is not much improvement in the accuracy of the calculation when increasing the size of JKFIT from cc-pVDZ to cc-pVTZ, but on increase from cc-pVTZ to cc-pVQZ, significant improvement in the accuracy is seen. A JKFIT of at least cc-pVQZ is required to obtain maximum errors smaller than 0.0050 kcal mol$^{-1}$, which can be considered acceptably small for interaction energies. By cc-pV5Z JKFIT basis, errors compared to the non-density-fitted result are negligibly small with a maximum error of 0.0031 kcal mol$^{-1}$ and a mean absolute error of 0.0005 kcal mol$^{-1}$. 

\begin{table}
	\centering
	\caption{Error statistics in kcal mol$^{-1}$ for density fitted SCF with different  JKFIT bases with reference to non-density-fitted SCF/aug-cc-pVDZ interaction energies. MAE is the mean absolute error, RMSE is the root mean squared error and MAXE is the maximum absolute error.}
	\label{df-hf-err}
	\begin{tabular}{cccc}
		\hline
		JKFIT & MAE & RMSE & MAXE \\
		\hline
		cc-pVDZ &0.0049  & 0.0066 & 0.0178\\
		cc-pVTZ& 0.0042 & 0.0054&  0.0134\\
		cc-pVQZ & 0.0015  & 0.0021 & 0.0050 \\
		cc-pV5Z & 0.0005 & 0.0009 &  0.0031\\
		\hline
	\end{tabular}
	\end{table}
	
\subsubsection{CCSD(T) correlation interaction energy}

\begin{figure}
		\centering
		\includegraphics*[width=0.45\textwidth]{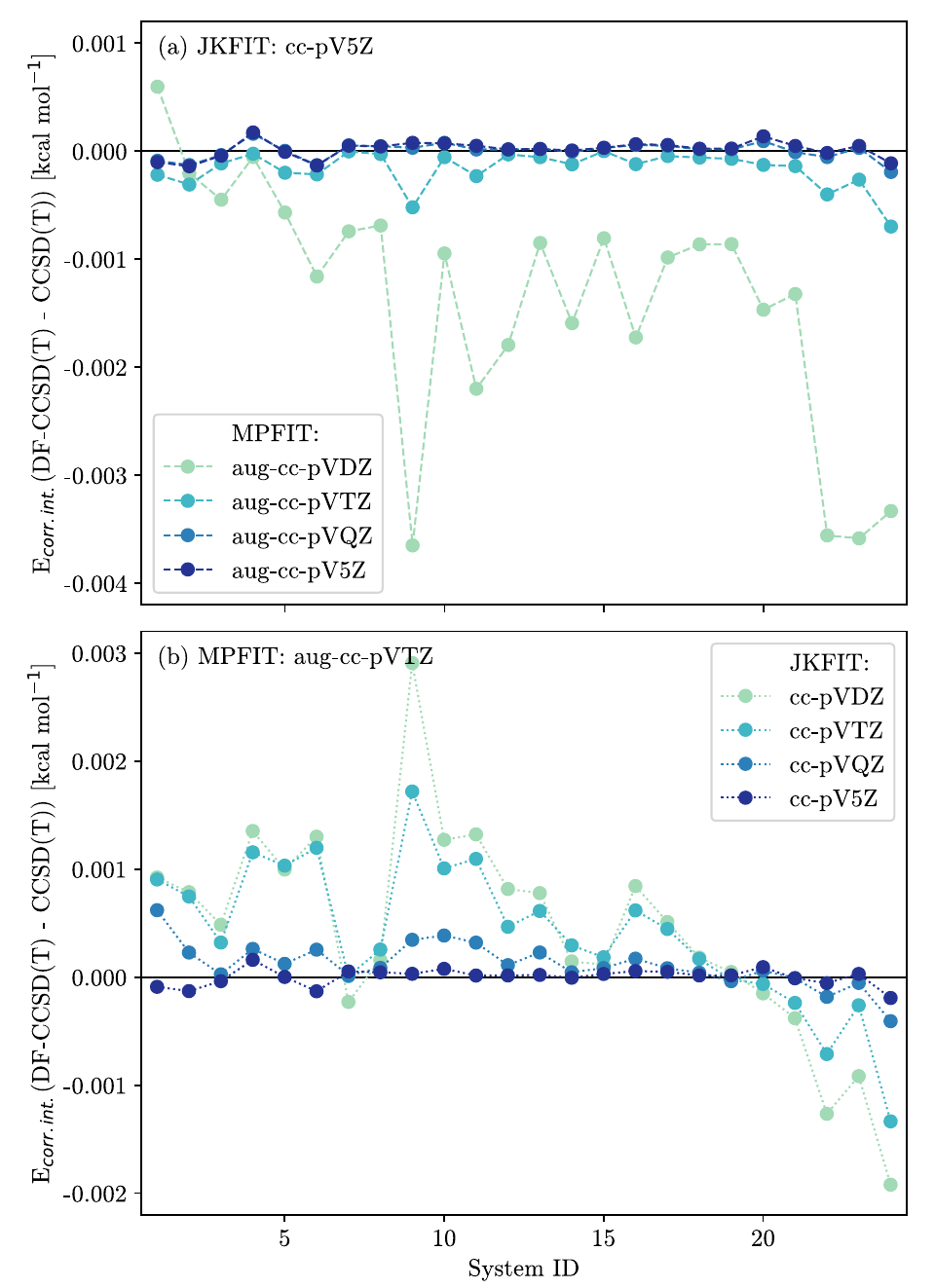} 
		\caption{Error resulting from density fitting compared to the non-density-fitted CCSD(T)/aug-cc-pVDZ correlation interaction energies. Panel (a) varies the MPFIT basis set, keeping JKFIT at cc-pV5Z; panel (b) varies the JKFIT basis (at the SCF step of the calcualtions), while keeping MPFIT at aug-cc-pVTZ. }
		\label{df-corr}
	\end{figure}

Using a cc-pV5Z JKFIT basis set for the Hartree-Fock part of the calculation, the accuracy of the MPFIT density fitting basis sets is tested with a reference to the non-density-fitted CCSD(T)/aug-cc-pVDZ results. Figure \ref{df-corr}a and Table \ref{df-corr-err} demonstrate that already at the aug-cc-pVTZ MPFIT basis set, all measures of the error introduced by the density fitting are consistently lower than 0.001 kcal mol$^{-1}$. As such, we ascertain that for aug-cc-pVDZ calculations  a MPFIT of aug-cc-pVTZ is sufficiently accurate to obtain correlation interactions with a high degree of fidelity to the reference result. 

Finally, using the aug-cc-pVTZ basis set for the MPFIT, we iterate through the series of JKFIT basis sets once again, this time comparing to the conventional CCSD(T) correlation interaction energy. Figure \ref{df-corr}b and Table \ref{df-corr-err} demonstrates that the combination of MPFIT = aug-cc-pVTZ with JKFIT = cc-pVQZ results in a maximum error of less than 0.001 kcal mol$^{-1}$ for correlation interaction energies across the A24 dataset. 
We use the aforementioned combination of auxiliary basis sets as the error introduced into the calculation from the density fitting is negligible at the scale of the method error. 

\begin{table}
	\centering
	\caption{Error statistics in kcal mol$^{-1}$ for the MPFIT and JKFIT basis sets with reference to the conventional CCSD(T)/aug-cc-pVDZ correlation interaction energies. MAE is the mean absolute error, RMSE is the root mean squared error and MAXE is the maximum absolute error.}
	\label{df-corr-err}
	\begin{tabular}{cccc}
		\hline\hline
	\multicolumn{4}{c}{JKFIT: cc-pV5Z} \\
		\hline
		MPFIT & MAE & RMSE & MAXE \\
		\hline
	aug-cc-pVDZ &0.0014 & 0.0018& 0.0036\\
		aug-cc-pVTZ&0.0002 & 0.0002 & 0.0007\\
		aug-cc-pVQZ &0.0001  &0.0001 & 0.0002\\
		aug-cc-pV5Z & 0.0001 &0.0001  & 0.0002\\
		\hline\hline
		\multicolumn{4}{c}{MPFIT: aug-cc-pVTZ} \\
		\hline
        JKFIT & MAE & RMSE & MAXE \\
		\hline
		cc-pVDZ & 0.0008& 0.0011& 0.0029\\
		cc-pVTZ & 0.0006& 0.0008& 0.0017\\
		cc-pVQZ &0.0002 & 0.0002&0.0006\\
		cc-pV5Z &0.0001& 0.0001& 0.0002\\
		\hline\hline
	\end{tabular}
\end{table}

We note that, generally, the errors introduced by density fitting in the CCSD(T) correlation interaction are much smaller than those introduced as a result of density fitting in the SCF interaction energy. 
Thus, the density fitting can be employed with minimal error introduction in the correlated component of calculations, while the already low cost SCF calculation could be carried out without density fitting. 
Furthermore, we determine that a density fitting basis at the same cardinal number as the atomic orbital basis is sufficient to obtain accurate total and interaction energies, i.e., the density fitting errors in correlation energies are approximately one order of magnitude
smaller than the method errors (Section \ref{dc-ccsdt}). 
Here, however, we utilize density fitting basis sets that are one (for MPFIT) or two (for JKFIT) cardinal numbers larger than required to further mitigate the impact of the density fitting onto the accuracy of the SVD methods,
and to be able to go to larger atomic orbital basis set sizes (Section \ref{larger_bs}) without changing the density fitting parameters. 


\subsection{Benchmarking DC-CCSDT to CCSDT(Q)}\label{dc-ccsdt}

\begin{figure}
	\centering
	\includegraphics*[width=0.45\textwidth]{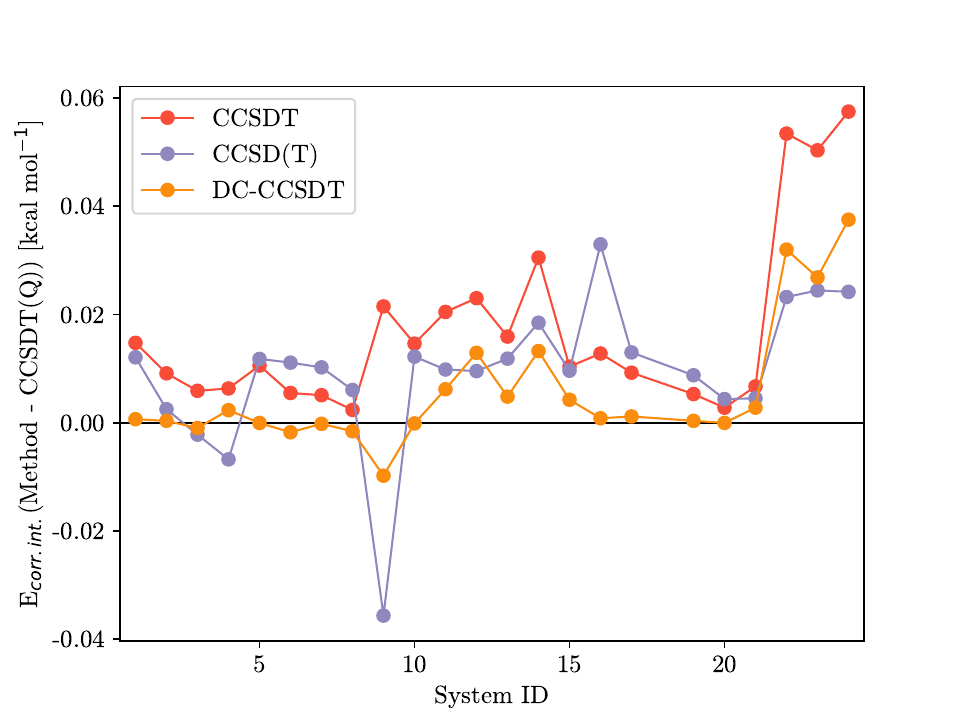}
	\caption{Performance of higher order methods compared to the CCSDT(Q) correlation interaction energy. All calculations are carried out in an aug-cc-pVDZ atomic orbital basis set.}
	\label{eint}
\end{figure}

\begin{table*}
\centering
\caption{Total correlation interaction energy and associated error relative to CCSDT(Q) for each system in the A24 dataset. All calculations are carried out in the aug-cc-pVDZ atomic orbital basis set, with counterpoise corrections. MAE is the mean absolute error, RMSE is the root mean squared error and MAXE is the maximum absolute error.}
	\label{error-stats}
\begin{tabular}{clccccccc}
\hline
ID	&	Label	&	\multicolumn{4}{c}{$E_{\text{corr. int.}}$ [kcal mol$^{-1}$]} & \multicolumn{3}{c}{$E_{\text{corr. int.}}$ (Method - CCSDT(Q)) [kcal mol$^{-1}$]}\\
\cmidrule(lr){3-6} 
 \cmidrule(lr){7-9} 
 & & CCSDT(Q)\footnote{Ref. \citenum{rezac:2013b}}	&	CCSDT	&	CCSD(T)\footnote{Ref. \citenum{burns:2014}}&	DC-CCSDT		&		$\delta$CCSDT	&	$\delta$CCSD(T)	&	$\delta$DC-CCSDT	\\
\hline
1	&	water - ammonia	&	-1.2012	&	-1.1865	&	-1.1892	&	-1.2006		&		0.0148	&	0.0121	&	0.0007	\\
2	&	water dimer	&	-0.7361	&	-0.7270	&	-0.7336	&	-0.7358		&		0.0091	&	0.0025	&	0.0003	\\
3	&	HCN dimer	&	-0.2469	&	-0.2410	&	-0.2491	&	-0.2478		&		0.0059	&	-0.0022	&	-0.0010	\\
4	&	HF dimer	&	-0.2145	&	-0.2082	&	-0.2213	&	-0.2122		&		0.0063	&	-0.0068	&	0.0023	\\
5	&	ammonia dimer	&	-1.2245	&	-1.2139	&	-1.2127	&	-1.2245		&		0.0106	&	0.0118	&	-0.0001	\\
6	&	HF - methane	&	-0.9453	&	-0.9398	&	-0.9342	&	-0.9470		&		0.0055	&	0.0111	&	-0.0018	\\
7	&	ammonia methane	&	-0.6442	&	-0.6392	&	-0.6340	&	-0.6444		&		0.0051	&	0.0102	&	-0.0002	\\
8	&	water - methane	&	-0.5365	&	-0.5342	&	-0.5305	&	-0.5381		&		0.0024	&	0.0061	&	-0.0016	\\
9	&	formaldehyde dimer	&	-1.1048	&	-1.0833	&	-1.1405	&	-1.1146		&		0.0215	&	-0.0356	&	-0.0098	\\
10	&	water - ethene	&	-1.2133	&	-1.1987	&	-1.2011	&	-1.2134		&		0.0146	&	0.0122	&	-0.0001	\\
11	&	formaldehyde - ethene	&	-1.2353	&	-1.2149	&	-1.2255	&	-1.2291		&		0.0205	&	0.0098	&	0.0062	\\
12	&	ethyne dimer	&	-0.7458	&	-0.7228	&	-0.7363	&	-0.7329		&		0.0230	&	0.0095	&	0.0129	\\
13	&	ammonia - ethene	&	-1.1053	&	-1.0893	&	-1.0934	&	-1.1004		&		0.0159	&	0.0118	&	0.0048	\\
14	&	ethene dimer	&	-1.7653	&	-1.7348	&	-1.7468	&	-1.7521		&		0.0305	&	0.0185	&	0.0132	\\
15	&	methane - ethene	&	-0.7657	&	-0.7554	&	-0.7561	&	-0.7615		&		0.0104	&	0.0096	&	0.0042	\\
16	&	borane - methane	&	-2.1842	&	-2.1714	&	-2.1512	&	-2.1834		&		0.0128	&	0.0330	&	0.0008	\\
17	&	methane - ethane	&	-1.3848	&	-1.3755	&	-1.3718	&	-1.3836		&		0.0092	&	0.0130	&	0.0012	\\
18	&	methane - ethane	&	-	&	-	&	-	&	-		&		-	&	-	&	-	\\
19	&	methane dimer	&	-0.8933	&	-0.8880	&	-0.8845	&	-0.8930		&		0.0053	&	0.0088	&	0.0004	\\
20	&	Ar - methane	&	-0.5902	&	-0.5874	&	-0.5859	&	-0.5903		&		0.0028	&	0.0044	&	0.0000	\\
21	&	Ar - ethene	&	-0.6140	&	-0.6073	&	-0.6095	&	-0.6112		&		0.0067	&	0.0045	&	0.0028	\\
22	&	ethene - ethyne	&	-2.3032	&	-2.2498	&	-2.2800	&	-2.2712		&		0.0534	&	0.0232	&	0.0320	\\
23	&	ethene dimer	&	-2.6930	&	-2.6426	&	-2.6685	&	-2.6661		&		0.0503	&	0.0244	&	0.0268	\\
24	&	ethyne dimer	&	-2.0675	&	-2.0100	&	-2.0433	&	-2.0300		&		0.0575	&	0.0242	&	0.0375	\\
\hline
& MAE & & & & & 0.0164 & 0.0127 & 0.0067 \\
& RMSE & & & & & 0.0233 & 0.0160 & 0.0127 \\
& MAXE & & & & & 0.0575 & 0.0356 & 0.0375\\ 
\hline
\end{tabular}
\end{table*}

As can be seen from Fig. \ref{eint} and Table \ref{error-stats}, the DC-CCSDT results follow the  reference  very closely and, in the majority of cases, outperforms both CCSD(T) and CCSDT.  The error statistics in Table \ref{error-stats} demonstrate that DC-CCSDT has, on average, the best performance, with the lowest mean absolute errors and root mean square errors. Importantly, we demonstrate that CCSDT is the least reliable method for reproducing CCSDT(Q) interaction energies, of those considered here. This is in line with previous studies that show that CCSDT is not so accurate at reproducing CCSDT(Q) non covalent interactions compared to other CC methods.\cite{rezac:2013, semidalas:2025, karton:2021} Furthermore, compared to the non-perturbative CCSDTQ method using the aug-cc-pVDZ basis sets, CCSD(T) and CCSDT have similar results for small dispersion and hydrogen bound interaction energies,\cite{rezac:2013} while for increasingly large Pariser-Parr-Pople\cite{pariser:1953a, pariser:1953b, pople:1953} model dispersion interactions, CCSD(T) routinely outperforms CCSDT, compared to CCSDTQ.\cite{lambie:2024} However, the comparatively poor performance of CCSDT compared to higher order CC methods is somewhat troublesome, as fixed node diffusion Monte Carlo results have been shown to agree better with (approximate) CCSDT\cite{schafer:2024, shi:2025} results compared to CCSD(T) calculations, thus reiterating the need for detailed studies teasing out all possible sources of error within both CC, beyond CCSDT, and fixed-node diffusion Monte Carlo results to fully understand where the discrepancies between these two state-of-the-art methods are originating from. 

The performance of all methods deteriorates somewhat for systems 22-24. These three geometries are included in the A24 dataset to be representative of $\pi-\pi$ stacking interactions. The actual reason of this worsening is not clear, but it is worth noting that exactly these three structures are shifted away from the minima on the potential energy surface  (these dimers have been taken in the sandwich configuration with the 3.5 \AA{} separation, which is the sum of the van der Waals radii of the carbon atoms\cite{rezac:2011b}). Furthermore, with the aug-cc-pVDZ basis they even have an overall positive interaction energy, i.e. sit on the repulsive wall of the intermolecular interaction potential. 


Nevertheless, the benchmarking carried out up to this point clearly demonstrates that DC-CCSDT can 
be used as a reference methodology for obtaining highly accurate non-covalent interaction energies approaching CCSDT(Q) quality.

\subsection{Benchmarking SVD-DC-CCSDT and derivatives}

\begin{figure}[H]
	\centering
\includegraphics*[width=0.45\textwidth]{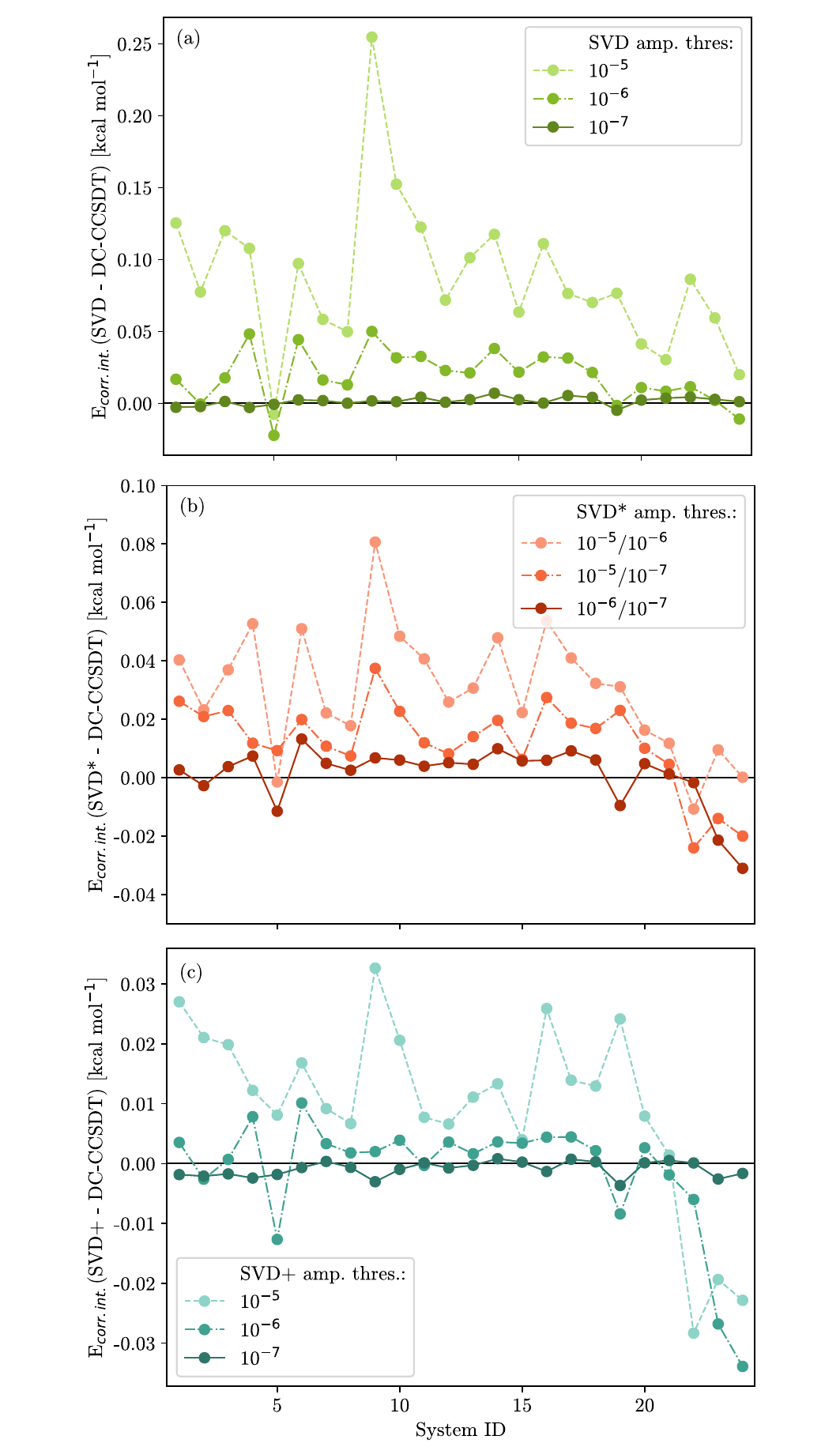}
	\caption{Performance of tensor-decomposed methods, (a) SVD-DC-CCSDT, (b) SVD-DC-CCSDT* and (c) SVD-DC-CCSDT+, for non-covalent interactions of the A24 dataset. The atomic orbital basis set is aug-cc-pVDZ, while for density fitting MPFIT of aug-cc-pVTZ and JKFIT of cc-pVQZ are used. Note: the different \textit{y}-scales on the plots.}
	\label{rel_to_dc_ccsdt}
\end{figure}

Next, the performance of the SVD family of methods, including SVD-DC-CCSDT (Figure \ref{rel_to_dc_ccsdt}a), SVD-DC-CCSDT* (Figure \ref{rel_to_dc_ccsdt}b) and SVD-DC-CCSDT+ (Figure \ref{rel_to_dc_ccsdt}c), is examined, with reference to DC-CCSDT correlation interaction energies. 
From Figure \ref{rel_to_dc_ccsdt}, it is immediately apparent that the (T)-corrected implementation, SVD-DC-CCSDT+, provides the most accurate results when benchmarked to the non-SVD reference. With the SVD amplitude threshold increased to $10^{-7}$, or in the case of the SVD-DC-CCSDT* method, to $10^{-6}/10^{-7}$, all the SVD methods provide excellent agreement to DC-CCSDT with sub-0.01 kcal mol$^{-1}$ errors (see Table \ref{svd_err}). At the looser SVD amplitude thresholds of $10^{-6}$ (and even to some extent $10^{-5}$), SVD-DC-CCSDT+ also performs excellently. In contrast to this, the error associated with the SVD-DC-CCSDT interaction has increased by almost one order of magnitude, to a MAE of 0.0220 kcal mol$^{-1}$. For the SVD-DC-CCSDT* method, when the amplitude correction is loosened to $10^{-5}/10^{-7}$ or further to $10^{-5}/10^{-6}$, the mean absolute error is comparable to the SVD-DC-CCSDT error with a threshold $10^{-6}$.

In Table \ref{svd_to_ccsdtq}, we compare the performance of the SVD methods directly to CCSDT(Q). As expected, the performance of SVD-DC-CCSDT+ with an SVD amplitude threshold of 10$^{-6}$ is comparable to DC-CCSDT itself. This therefore suggests that the (T)-corrected, SVD-DC-CCSDT+ method with the 10$^{-6}$ threshold can be seen as a low-cost alternative to CCSDT(Q) for calculation of interaction energies.
Total correlation interaction energies for the SVD methods across different SVD amplitude thresholds, analogous to those reported in Table \ref{error-stats}, are documented in Table \ref{svd_corr_int}. 

\begin{table}
	\centering
	\caption{Error statistics in kcal mol$^{-1}$ for the SVD-DC-CCSDT-based approaches relative to DC-CCSDT.}
	\label{svd_err}
	\begin{tabular}{cccc}
		\hline
		SVD amp. thres. & MAE & RMSE & MAXE \\
		\hline
		\multicolumn{4}{c}{SVD-DC-CCSDT} \\
        \hline
		10$^{-5}$ &0.0874& 0.1005&0.2547\\
		10$^{-6}$& 0.0220 &0.0260 & 0.0500\\
		10$^{-7}$ &  0.0026&0.0031 & 0.0071\\
		\hline			\multicolumn{4}{c}{SVD-DC-CCSDT*} \\
		\hline
10$^{-5}$/10$^{-6}$ & 0.0312 & 0.0364 & 0.0806 \\
10$^{-5}$/10$^{-7}$ & 0.0170 & 0.0187 & 0.0374 \\
10$^{-6}$/10$^{-7}$ & 0.0076 & 0.0100 & 0.0311 \\
\hline
			\multicolumn{4}{c}{SVD-DC-CCSDT+} \\
		\hline
		10$^{-5}$ & 0.0156& 0.0177& 0.0327 \\
		10$^{-6}$& 0.0063 &0.0101 & 0.0339\\
		10$^{-7}$ & 0.0012 & 0.0016 & 0.0037\\
		\hline
	\end{tabular}
\end{table}

\begin{table}
	\centering
	\caption{Error statistics in kcal mol$^{-1}$ for the triples-truncated coupled-cluster-based methods relative to CCSDT(Q).}
	\label{svd_to_ccsdtq}
	\begin{tabular}{ccccc}
		\hline
			Method & SVD amp. thres. & MAE & RMSE & MAXE \\
			\hline
        CCSDT &- & 0.0164 & 0.0233 & 0.0575 \\
        CCSD(T) & - & 0.0127 & 0.0160 & 0.0356\\
		 DC-CCSDT & -& 0.0067 & 0.0127 & 0.0375 \\
		 \hline
	SVD-DC-CCSDT & 10$^{-5}$ & 0.0900 &0.1052 & 0.2449\\
		  & 10$^{-6}$&0.0256 & 0.0303& 0.0514\\
		& 10$^{-7}$ & 0.0084 & 0.0143& 0.0387\\
		\hline
    SVD-DC-CCSDT* &10$^{-5}$/10$^{-6}$ & 0.0344 & 0.0395 & 0.0709 \\
    &10$^{-5}$/10$^{-7}$ & 0.0169 & 0.0191 & 0.0328 \\
    & 10$^{-6}$/10$^{-7}$ & 0.0085 & 0.0112 & 0.0303 \\
    \hline
		SVD-DC-CCSDT+ & 10$^{-5}$ & 0.0146&0.0170 & 0.0277 \\
		 & 10$^{-6}$& 0.0066 & 0.0092& 0.0260\\
		 & 10$^{-7}$ & 0.0071& 0.0124 & 0.0359\\
		\hline
	\end{tabular}
\end{table}

We highlight a very good performance of the SVD-DC-CCSDT* method when the SVD-(T) correction is calculated using $10^{-6}/10^{-7}$. Crucially, this method is performed entirely within the SVD construction and removes the necessity for the only ($\mathcal{O}$(\textit{N}$^7$)) step in a  SVD-DC-CCSDT+ calculation: evaluation of the non-SVD (T)-energy. This ultimately means that results approaching CCSDT(Q) quality can be obtained within CCSD scaling. The agreement to both DC-CCSDT (Table \ref{svd_err}) and CCSDT(Q) (Table \ref{svd_to_ccsdtq}) is remarkable, with mean absolute errors of less than 0.01 kcal mol$^{-1}$ when $10^{-6}/10^{-7}$ is chosen for the SVD-DC-CCSDT* method, outperforming both CCSD(T) and CCSDT. 

\subsection{Computational cost}\label{comp-time}

In this section, we compare the time required to calculate the complete interaction energy (monomer \textit{a} with counterpoise corrections, monomer \textit{b} with counterpoise corrections and the dimer) within the SVD implementation (containing both SVD-DC-CCSDT and SVD-DC-CCSDT+ results), compared to the full DC-CCSDT result in the aug-cc-pVDZ atomic orbital basis set. We report the timings for four systems; systems 4 (HF dimer), 15 (methane-ethene dimer), 16 (borane-methane dimer), and 23 (ethene dimer) (Table \ref{timings}). These four systems represent the worst speed up offered by SVD, the best speed up offered by SVD, the smallest number of correlated electrons and the largest number of correlated electrons, respectively. The timings of all systems can be found in Table \ref{supp_timings}.

\begin{table}[H]
	\centering
	\caption{Timings of SVD-DC-CCSDT calculations, compared to DC-CCSDT calculations in seconds in the aug-cc-pVDZ.}
	\label{timings}
	\begin{tabular}{ccccccc}
		\hline
		System & N$_{\text{corr.elect.}}$ & N$_{\text{orb.}}$& DC-CCSDT & \multicolumn{3}{c}{SVD amp. thres.} \\
				\cmidrule(lr){5-7} 
		& & &  & $10^{-5}$ & $10^{-6}$ & $10^{-7}$ \\
		\hline
	\multicolumn{7}{c}{Total timings [s]}\\
	\hline
		4 &	16&  62  &674	&15 &	31	&224 \\
		15& 20&	138&66547&	296	&1642	&4672\\ 
		16& 14& 107	&3425	&47	&261&	796\\ 
		23& 24 &	160&62546	&396&2972		&26098 \\
		\hline
			\multicolumn{7}{c}{Speed up factors}\\
		\hline
		 4&	16& 62& 1&	46	&22	&3\\
		15&	20& 138& 1&	225&	41	&14 \\
		16&	14& 107& 1&	73	&13&	4\\
		23&	24& 160& 1&	158	&21&	2\\
	\hline
	\end{tabular}
\end{table}

As can be seen in Table \ref{timings}, the SVD implementation of DC-CCSDT with the SVD amplitude thresholds of $10^{-5}$, provides a speed up over the DC-CCSDT calculation of between 46 and 225 times. However, as was previously discussed, for the optimal accuracy the SVD-DC-CCSDT+ with an amplitude threshold of $10^{-6}$ is recommended, which still offers speed ups over conventional DC-CCSDT of between 13 and 41 times. 
Yet, the deviation of SVD-DC-CCSDT+ with this threshold from CCSDT(Q) is not higher than 0.01 kcal mol$^{-1}$. 

We have to mention that the reference DC-CCSDT calculations have been carried out using an automatically generated code from the Quantwo\cite{quantwo} package. If the DC-CCSDT code was optimized, the efficiency of the DC-CCSDT calculations could be higher. However, we do not expect this to change the principle gains in the efficiency due to the SVD approximation. Furthermore, for bigger systems the speed up is expected to grow due to the higher scaling of the conventional DC-CCSDT method with system size.
In addition, the size of the MPFIT basis directly affects the timings of SVD methods\cite{rickert:2025} and is negligible for DC-CCSDT,
which uses four-index integrals in residual calculation. Therefore, the speed up would be substantially greater if an aug-cc-pVDZ MPFIT basis were employed. 

\subsection{Convergence of post CCSD(T) corrections with basis set size}\label{larger_bs}

Finally, having established the high accuracy of SVD-DC-CCSDT+, especially so with the SVD amplitude threshold of $10^{-6}$, we focus on the post CCSD(T) corrections for the interaction energies, also aiming at converging it with the basis set size. 
For that we use the aug-cc-pVTZ atomic orbital basis set, which is unprecedented for this level of theory. 
We believe the availability of such low-cost high-level methods can make basis set converged calculation of post-CCSD(T) correction for moderately-sized systems (i.e. up to 30-40 correlated electrons) more routine.  

\begin{figure}
	\centering
	\includegraphics[width=0.45\textwidth]{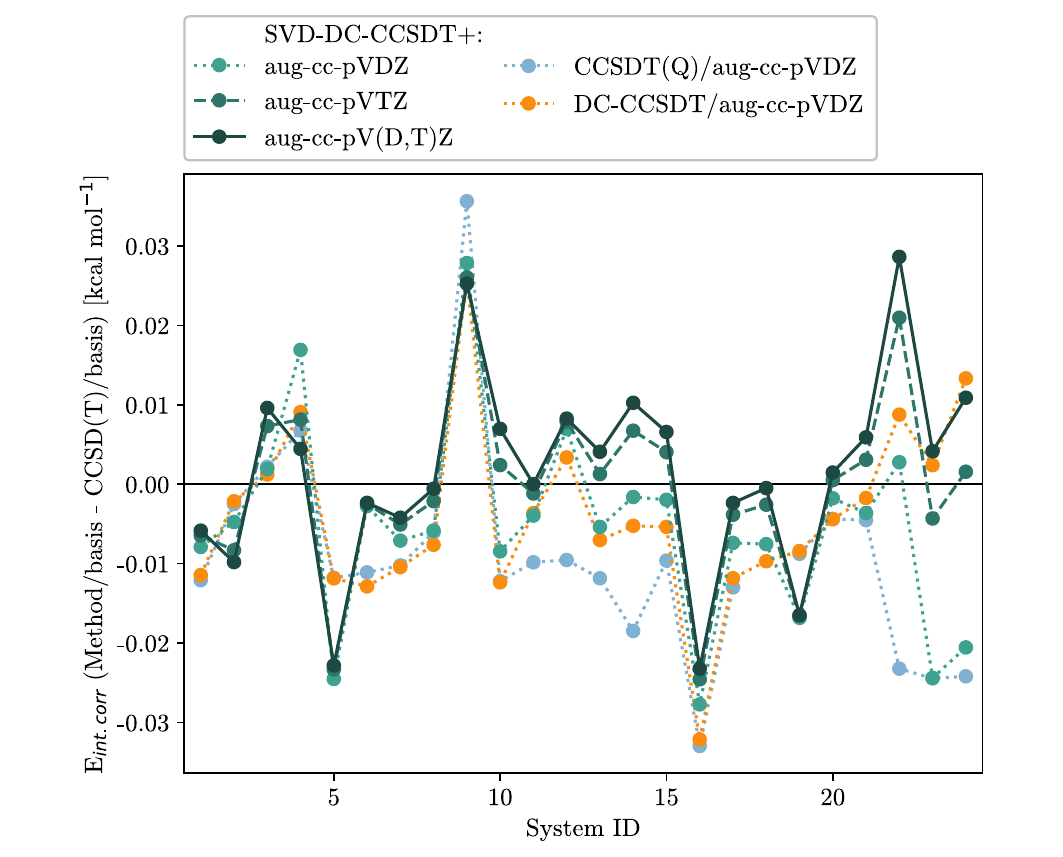}
	\caption{Post CCSD(T) corrections, with reference to the CCSD(T) correlation interaction energy in the corresponding atomic basis set size. An SVD threshold of 10$^{-6}$ is used.}
	\label{corrections}
\end{figure}

Comparison of the SVD results obtained in the aug-cc-pVTZ atomic orbital basis set to those calculated in the aug-cc-pVDZ atomic orbital basis set (Figure \ref{corrections}, Table \ref{post-ccsdt-corr}) provides a proxy for how converged the post CCSD(T) correction is with respect to basis set size. 
Generally, we find that the SVD-DC-CCSDT+ correction obtained in the aug-cc-pVDZ atomic orbital basis is in good agreement with the correction obtained in the aug-cc-pVTZ atomic orbital basis set. There are, however, a few exceptions, the most notable of which are systems 22-24 where the difference between the SVD-DC-CCSDT+ correction in the aug-cc-pVTZ and aug-cc-pVDZ ranges from 0.0182 to 0.0221 kcal mol$^{-1}$. Across the complete A24 dataset, however, the agreement between the two SVD-DC-CCSDT+ basis set sizes is generally very good, with a mean absolute error of 0.0040 kcal mol$^{-1}$. 
Although for most of the studied systems a small basis set (aug-cc-pVDZ) is a reasonable approximation for the post CCSD(T) corrections, a possibility to go to the larger basis sets is very desirable to probe the basis set convergence. Using the aug-cc-pVDZ and aug-cc-pVTZ results, we can then extrapolate to the complete basis set limit\cite{helgaker:2000, halkier:1999} and obtain post CCSD(T) corrections approaching basis set convergences.
As concerns the computational cost, the SVD-DC-CCSDT+/aug-cc-pVTZ calculations are, on average, four times more costly (in CPU time) than the SVD-DC-CCSDT+/aug-cc-pVDZ calculations (note that the fitting basis sets were kept the same, see section \ref{df-testing}). The SVD-DC-CCSDT+ correlation energies in aug-cc-pVDZ and aug-cc-pVTZ basis sets calculated with 10$^{-6}$ and 10$^{-7}$ SVD amplitude thresholds are reported in full in Tables \ref{1} - \ref{4} and can be used as reference correlation interaction energies for future benchmarking studies.

\section{Conclusions}

In conclusion, benchmarking of the DC approach, and its SVD counterpart, to platinum standard CCSDT(Q) correlation interaction energies is carried out for the A24 dataset. DC-CCSDT is shown to outperform the conventional CCSDT and CCSD(T). Furthermore, SVD-DC-CCSDT+ with an SVD amplitude threshold of $10^{-6}$ is shown to perform excellently when benchmarked to CCSDT(Q) and can be used as a low-cost benchmark method approaching CCSDT(Q) quality with mean absolute errors of less than 0.01 kcal mol$^{-1}$. Having access to such a low-cost, high accuracy method allows us to obtain high quality correlation interaction energies in atomic basis set sizes of unprecedented size and will, in future, allow calculation of larger systems sizes at a post CCSD(T) level of theory. We also demonstrate a proof-of-concept for the even lower scaling $\mathcal{O}$(\textit{N}$^6$) SVD-DC-CCSDT* method, which also outperforms both CCSD(T) and CCSDT when the sufficiently tight SVD amplitude thresholds are specified. Finally, we provide correlation energies for the complete A24 dataset using the SVD-DC-CCSDT+ (SVD amplitude thresholds of 10$^{-6}$ and 10$^{-7}$) in the aug-cc-pVDZ and aug-cc-pVTZ atomic orbital basis sets, which will be invaluable for future benchmarking studies.

 \section*{Supplementary Material}
 
 See the supplementary material for correlation interaction energies for all DC-CCSDT and SVD methods discussed, the timings required for SVD-DC-CCSDT calculations, post CCSD(T) corrections for higher order methods and correlation energies for the complete A24 dataset using SVD-DC-CCSDT+ in the aug-cc-pVDZ and aug-cc-pVTZ atomic orbital basis set sizes. 
 
 \begin{acknowledgments}
 	Financial support from the Max-Planck Society is gratefully acknowledged.
    In addition, C.R. thanks Studienstiftung des deutschen Volkes for a Masters scholarship.

\end{acknowledgments}

\section*{Author Declaractions}
\subsection*{Conflicts of Interest}
The authors have no conflicts of interest to declare.

\subsection*{Author contributions}
\textbf{S. Lambie}: Data curation (lead), investigation (lead), writing - original draft preparation (lead), review and editing (equal). 
\textbf{C. Rickert}: Methodology (lead), writing - review and editing (supporting).
\textbf{D. Usyvat}: Writing - review and editing (supporting), supervision (supporting). 
\textbf{A. Alavi:} Conceptualization (equal), writing - review and editing (equal), supervision (equal).
\textbf{D. Kats:} Conceptualization (equal), writing - review and editing (equal), supervision (equal).

\section*{Data Availability Statement}

The data that support the findings of this study are available within the article and its supplementary material.

\bibliography{mpi-fkf}

\end{document}


\maketitle

    \section{Correlation interaction energies}

\begin{table}[H]
        \caption{Correlation interaction energies for the SVD methods at a variety of different SVD amplitude thresholds for the A24 dataset. All values reported in kcal mol$^{-1}$ and calculated in the aug-cc-pVDZ atomic orbital basis set with counterpoise corrections.}
    \begin{sideways}
    \begin{minipage}{.90\textheight}
        \resizebox{.90\textheight}{!}{%
            \setstretch{1.0}
        \label{svd_corr_int}
        \begin{tabular}{clccccccccccc}
        \hline
           ID	&	Label	&	CCSDT(Q)	&	DC-CCSDT 	& 	\multicolumn{3}{c}{SVD-DC-CCSDT} &	\multicolumn{3}{c}{SVD-DC-CCSDT*} &\multicolumn{3}{c}{SVD-DC-CCSDT+}	\\	
           \cmidrule(lr){5-7}
           \cmidrule(lr){8-10}
           \cmidrule(lr){11-13}					
	&		&		&		&	10$^{-5}$	&	10$^{-6}$	&	10$^{-7}$	&	10$^{-5}$/10$^{-6}$	&	10$^{-5}$/10$^{-7}$	&	10$^{-6}$/10$^{-5}$	&	10$^{-5}$	&	10$^{-6}$	&	10$^{-7}$	\\
    \hline
1	&	water - ammonia	&	-1.2012	&	-1.2006	& 	-1.0752	&	-1.1838	&	-1.2033	&	-1.1603	&	-1.1745	&	-1.1980	&	-1.1736	&	-1.1971	&	-1.2024	\\
2	&	water dimer	&	-0.7361	&	-0.7358	& 	-0.6584	&	-0.7362	&	-0.7380	&	-0.7126	&	-0.7149	&	-0.7385	&	-0.7147	&	-0.7384	&	-0.7379	\\
3	&	HCN dimer	&	-0.2469	&	-0.2478	& 	-0.1278	&	-0.2301	&	-0.2465	&	-0.2109	&	-0.2249	&	-0.2441	&	-0.2280	&	-0.2471	&	-0.2496	\\
4	&	HF dimer	&	-0.2145	&	-0.2122	& 	-0.1045	&	-0.1639	&	-0.2151	&	-0.1596	&	-0.2005	&	-0.2048	&	-0.2000	&	-0.2044	&	-0.2146	\\
5	&	ammonia dimer	&	-1.2245	&	-1.2245	& 	-1.2321	&	-1.2468	&	-1.2252	&	-1.2260	&	-1.2153	&	-1.2360	&	-1.2164	&	-1.2372	&	-1.2263	\\
6	&	HF - methane	&	-0.9453	&	-0.9470	& 	-0.8498	&	-0.9028	&	-0.9446	&	-0.8961	&	-0.9272	&	-0.9338	&	-0.9302	&	-0.9369	&	-0.9477	\\
7	&	ammonia methane	&	-0.6442	&	-0.6445	& 	-0.5860	&	-0.6282	&	-0.6425	&	-0.6223	&	-0.6337	&	-0.6396	&	-0.6353	&	-0.6411	&	-0.6441	\\
8	&	water - methane	&	-0.5365	&	-0.5381	& 	-0.4883	&	-0.5252	&	-0.5380	&	-0.5203	&	-0.5307	&	-0.5356	&	-0.5314	&	-0.5363	&	-0.5387	\\
9	&	formaldehyde dimer	&	-1.1048	&	-1.1146	& 	-0.8599	&	-1.0646	&	-1.1129	&	-1.0340	&	-1.0772	&	-1.1079	&	-1.0820	&	-1.1126	&	-1.1176	\\
10	&	water - ethene	&	-1.2133	&	-1.2134	& 	-1.0611	&	-1.1817	&	-1.2123	&	-1.1650	&	-1.1908	&	-1.2074	&	-1.1928	&	-1.2095	&	-1.2144	\\
11	&	formaldehyde - ethene	&	-1.2353	&	-1.2291	& 	-1.1066	&	-1.1965	&	-1.2248	&	-1.1885	&	-1.2172	&	-1.2252	&	-1.2214	&	-1.2294	&	-1.2290	\\
12	&	ethyne dimer	&	-0.7458	&	-0.7329	& 	-0.6611	&	-0.7101	&	-0.7321	&	-0.7071	&	-0.7248	&	-0.7278	&	-0.7263	&	-0.7293	&	-0.7337	\\
13	&	ammonia - ethene	&	-1.1053	&	-1.1005	& 	-0.9992	&	-1.0792	&	-1.0978	&	-1.0698	&	-1.0864	&	-1.0959	&	-1.0893	&	-1.0988	&	-1.1007	\\
14	&	ethene dimer	&	-1.7653	&	-1.7521	& 	-1.6345	&	-1.7139	&	-1.7450	&	-1.7042	&	-1.7325	&	-1.7422	&	-1.7387	&	-1.7484	&	-1.7513	\\
15	&	methane - ethene	&	-0.7657	&	-0.7615	& 	-0.6981	&	-0.7398	&	-0.7590	&	-0.7392	&	-0.7552	&	-0.7558	&	-0.7575	&	-0.7581	&	-0.7613	\\
16	&	borane - methane	&	-2.1842	&	-2.1834	& 	-2.0723	&	-2.1511	&	-2.1832	&	-2.1296	&	-2.1559	&	-2.1774	&	-2.1575	&	-2.1789	&	-2.1847	\\
17	&	methane - ethane	&	-1.3848	&	-1.3836	& 	-1.3072	&	-1.3521	&	-1.3782	&	-1.3427	&	-1.3650	&	-1.3745	&	-1.3697	&	-1.3792	&	-1.3829	\\
18	&	methane - ethane	&	-	&	-1.0201	& 	-0.9500	&	-0.9986	&	-1.0160	&	-0.9878	&	-1.0033	&	-1.0141	&	-1.0072	&	-1.0180	&	-1.0198	\\
19	&	methane dimer	&	-0.8933	&	-0.8929	& 	-0.8164	&	-0.8944	&	-0.8978	&	-0.8619	&	-0.8700	&	-0.9025	&	-0.8688	&	-0.9014	&	-0.8966	\\
20	&	Ar - methane	&	-0.5902	&	-0.5902	& 	-0.5490	&	-0.5793	&	-0.5880	&	-0.5740	&	-0.5802	&	-0.5855	&	-0.5823	&	-0.5876	&	-0.5901	\\
21	&	Ar - ethene	&	-0.6140	&	-0.6112	& 	-0.5809	&	-0.6028	&	-0.6076	&	-0.5995	&	-0.6067	&	-0.6100	&	-0.6098	&	-0.6131	&	-0.6107	\\
22	&	ethene - ethyne	&	-2.3032	&	-2.2713	& 	-2.1850	&	-2.2597	&	-2.2669	&	-2.2820	&	-2.2953	&	-2.2730	&	-2.2996	&	-2.2772	&	-2.2711	\\
23	&	ethene dimer	&	-2.6930	&	-2.6661	& 	-2.6067	&	-2.6640	&	-2.6633	&	-2.6566	&	-2.6801	&	-2.6875	&	-2.6855	&	-2.6929	&	-2.6687	\\
24	&	ethyne dimer	&	-2.0675	&	-2.0300	& 	-2.0100	&	-2.0409	&	-2.0288	&	-2.0299	&	-2.0500	&	-2.0610	&	-2.0528	&	-2.0639	&	-2.0316	\\
\hline
        \end{tabular}
        }
        \end{minipage}
        \end{sideways}
    \end{table}

\section{Timings}

\begin{table}[H]
	\centering
	\caption{Total timings of all systems using SVD-DC-CCSDT compared to DC-CCSDT carried out in the aug-cc-pVDZ atomic orbital basis set. All timings reported are in seconds. }
    \label{supp_timings}
	\begin{tabular}{clcccccc}
		\hline
 ID& System & N$_{\text{corr. elec.}}$	& N$_{\text{orb.}}$ &DC-CCSDT& \multicolumn{3}{c}{SVD amp. thres} \\
				\cmidrule(lr){6-8} 
& &&&&10$^{-5}$ &	10$^{-6}$ & 10$^{-7}$\\
\hline
1	&	water - ammonia	&	16	&	89	&	3065	&	40	&	126	&	868	\\
2	&	water dimer	&	16	&	80	&	1518	&	23	&	92	&	724	\\
3	&	HCN dimer	&	20	&	106	&	11448	&	106	&	510	&	3533	\\
4	&	HF dimer	&	16	&	62	&	674	&	15	&	31	&	224	\\
5	&	ammonia dimer	&	16	&	98	&	3194	&	48	&	181	&	1450	\\
6	&	HF - methane	&	16	&	89	&	2194	&	31	&	89	&	384	\\
7	&	ammonia - methane	&	16	&	107	&	5590	&	55	&	214	&	841	\\
8	&	water - methane	&	16	&	98	&	2909	&	44	&	153	&	526	\\
9	&	formaldehyde dimer	&	24	&	124	&	27906	&	176	&	982	&	9058	\\
10	&	water - ethene	&	20	&	120	&	12126	&	119	&	591	&	3636	\\
11	&	formaldehyde - ethene	&	24	&	142	&	49769	&	267	&	1421	&	9866	\\
12	&	ethyne dimer	&	20	&	124	&	13842	&	162	&	1220	&	3047	\\
13	&	ammonia - ethene	&	20	&	129	&	21272	&	148	&	1126	&	4153	\\
14	&	ethene dimer	&	24	&	160	&	57033	&	372	&	3166	&	14639	\\
15	&	methane - ethene	&	20	&	138	&	66547	&	296	&	1642	&	4672	\\
16	&	borane - methane	&	14	&	107	&	3425	&	47	&	261	&	796	\\
17	&	methane - ethane	&	22	&	156	&	43631	&	275	&	1986	&	8738	\\
18	&	methane - ethane	&	22	&	156	&	47009	&	282	&	1858	&	8232	\\
19	&	methane dimer	&	16	&	116	&	4514	&	62	&	412	&	1912	\\
20	&	Ar - methane	&	16	&	80	&	1680	&	23	&	120	&	258	\\
21	&	Ar - ethene	&	20	&	102	&	6422	&	79	&	665	&	2723	\\
22	&	ethene - ethyne	&	22	&	142	&	38648	&	257	&	2160	&	9142	\\
23	&	ethene dimer	&	24	&	160	&	62546	&	396	&	2972	&	26098	\\
24	&	ethyne dimer	&	20	&	124	&	17813	&	158	&	1374	&	8206	\\
\hline
\end{tabular}
\end{table}

\section{Post CCSD(T) corrections}

\begin{table}[H]
    \centering
    \caption{Post CCSD(T) corrections, calculated by subtracting the CCSD(T) correlation interaction energy with counterpoise corrections in the corresponding basis set size. All values reported in kcal mol$^{-1}$. SVD thresholds were set to 10$^{-6}$.}
    \label{post-ccsdt-corr}
    \begin{tabular}{clcccc}
        \hline
 ID	&	Label	&  \multicolumn{3}{c}{aug-cc-pVDZ} & aug-cc-pVTZ \\
\cmidrule(lr){3-5} 
\cmidrule(lr){6-6}
& &  CCSDT(Q) 	&	DC-CCSDT &SVD-DC-CCSDT+	& 	SVD-DC-CCSDT+	\\
\hline
1	&	water - ammonia	&	-0.0121	&	-0.0114	&	-0.0079	& 	-0.0065	\\
2	&	water dimer	&	-0.0025	&	-0.0021	&	-0.0048	& 	-0.0083	\\
3	&	HCN dimer	&	0.0022	&	0.0012	&	0.0019	& 	0.0073	\\
4	&	HF dimer	&	0.0068	&	0.0091	&	0.0169	& 	0.0081	\\
5	&	ammonia dimer	&	-0.0118	&	-0.0118	&	-0.0245	& 	-0.0233	\\
6	&	HF - methane	&	-0.0111	&	-0.0129	&	-0.0027	& 	-0.0025	\\
7	&	ammonia methane	&	-0.0102	&	-0.0104	&	-0.0071	& 	-0.0051	\\
8	&	water - methane	&	-0.0061	&	-0.0076	&	-0.0058	& 	-0.0021	\\
9	&	formaldehyde dimer	&	0.0356	&	0.0259	&	0.0279	& 	0.0260	\\
10	&	water - ethene	&	-0.0122	&	-0.0123	&	-0.0084	& 	0.0024	\\
11	&	formaldehyde - ethene	&	-0.0098	&	-0.0036	&	-0.0039	& 	-0.0012	\\
12	&	ethyne dimer	&	-0.0095	&	0.0034	&	0.0070	& 	0.0079	\\
13	&	ammonia - ethene	&	-0.0118	&	-0.0070	&	-0.0054	& 	0.0013	\\
14	&	ethene dimer	&	-0.0185	&	-0.0052	&	-0.0016	& 	0.0067	\\
15	&	methane - ethene	&	-0.0096	&	-0.0054	&	-0.0020	& 	0.0041	\\
16	&	borane - methane	&	-0.0330	&	-0.0321	&	-0.0277	& 	-0.0246	\\
17	&	methane - ethane	&	-0.0130	&	-0.0118	&	-0.0074	& 	-0.0038	\\
18	&	methane - ethane	&	-	&	-0.0097	&	-0.0075	& 	-0.0026	\\
19	&	methane dimer	&	-0.0088	&	-0.0084	&	-0.0168	& 	-0.0166	\\
20	&	Ar - methane	&	-0.0044	&	-0.0044	&	-0.0018	& 	0.0005	\\
21	&	Ar - ethene	&	-0.0045	&	-0.0017	&	-0.0036	& 	0.0031	\\
22	&	ethene - ethyne	&	-0.0232	&	0.0088	&	0.0028	& 	0.0210	\\
23	&	ethene dimer	&	-0.0244	&	0.0024	&	-0.0244	& 	-0.0043	\\
24	&	ethyne dimer	&	-0.0242	&	0.0133	&	-0.0205	& 	0.0016	\\
\hline
    \end{tabular}
\end{table}

\clearpage

\section{Reference correlation energies using SVD-DC-CCSDT+}

\subsection{Atomic orbital basis set: aug-cc-pVDZ}

\subsubsection{SVD amplitude threshold 10$^{-6}$}

\begin{table}[H]
	\centering
	\caption{Correlation energies resulting from SVD-DC-CCSDT+ calculations, using an MPFIT of aug-cc-pVTZ, a JKFIT of cc-pVQZ, an atomic orbital basis set of aug-cc-pVDZ and an SVD amplitude threshold of 10$^{-6}$ with counterpoise corrections. All values in Hartree.}
	\label{1}
	\begin{tabular}{clccc}
		\hline
ID	&	System	&	Dimer	&	Monomer 1 	&	Monomer 2	\\
\hline
1	&	water - ammonia	&	-0.456895806	&	-0.233782641	&	-0.221205537	\\
2	&	water dimer	&	-0.468291735	&	-0.233475191	&	-0.233639863	\\
3	&	HCN dimer	&	-0.636579424	&	-0.317981910	&	-0.318203676	\\
4	&	HF dimer	&	-0.462711493	&	-0.231171854	&	-0.231213929	\\
5	&	ammonia dimer	&	-0.443594379	&	-0.220811404	&	-0.220811404	\\
6	&	HF - methane	&	-0.429406588	&	-0.196666836	&	-0.231246688	\\
7	&	ammonia methane	&	-0.418302700	&	-0.220796609	&	-0.196484413	\\
8	&	water - methane	&	-0.430468431	&	-0.233163771	&	-0.196450001	\\
9	&	formaldehyde dimer	&	-0.725082960	&	-0.361780680	&	-0.361529213	\\
10	&	water - ethene	&	-0.562574870	&	-0.327187474	&	-0.233459863	\\
11	&	formaldehyde - ethene	&	-0.689976484	&	-0.326909402	&	-0.361107838	\\
12	&	ethyne dimer	&	-0.588224364	&	-0.293765909	&	-0.293296178	\\
13	&	ammonia - ethene	&	-0.549456534	&	-0.326972648	&	-0.220732840	\\
14	&	ethene dimer	&	-0.656416064	&	-0.327053333	&	-0.326576422	\\
15	&	methane - ethene	&	-0.524370809	&	-0.326698417	&	-0.196464295	\\
16	&	borane - methane	&	-0.322198489	&	-0.196729960	&	-0.121996158	\\
17	&	methane - ethane	&	-0.560156972	&	-0.196674981	&	-0.361284146	\\
18	&	methane - ethane	&	-0.559471656	&	-0.196610886	&	-0.361238530	\\
19	&	methane dimer	&	-0.394510999	&	-0.196537296	&	-0.196537296	\\
20	&	Ar - methane	&	-0.366508389	&	-0.196349507	&	-0.169222450	\\
21	&	Ar - ethene	&	-0.496401559	&	-0.326296204	&	-0.169128347	\\
22	&	ethene - ethyne	&	-0.623508306	&	-0.326671753	&	-0.293207540	\\
23	&	ethene dimer	&	-0.657771712	&	-0.326740146	&	-0.326740146	\\
24	&	ethyne dimer	&	-0.589618074	&	-0.293164555	&	-0.293164555	\\
\hline
\end{tabular}
\end{table}

\clearpage

\subsubsection{SVD amplitude threshold 10$^{-7}$}

\begin{table}[H]
	\centering
	\caption{Correlation energies resulting from SVD-DC-CCSDT+ calculations, using an MPFIT of aug-cc-pVTZ, a JKFIT of cc-pVQZ, an atomic orbital basis set of aug-cc-pVDZ and an SVD amplitude threshold of 10$^{-7}$ with counterpoise corrections. All values in Hartree.}
	\label{2}
	\begin{tabular}{clccc}
		\hline
ID	&	System	&	Dimer	&	Monomer 1 	&	Monomer 2	\\
\hline
1	&	water - ammonia	&	-0.456802813	&	-0.233730864	&	-0.221155750	\\
2	&	water dimer	&	-0.468187171	&	-0.233424040	&	-0.233587259	\\
3	&	HCN dimer	&	-0.636517521	&	-0.317949624	&	-0.318170203	\\
4	&	HF dimer	&	-0.462604035	&	-0.231110946	&	-0.231151055	\\
5	&	ammonia dimer	&	-0.443479522	&	-0.220762625	&	-0.220762625	\\
6	&	HF - methane	&	-0.429326926	&	-0.196633733	&	-0.231182936	\\
7	&	ammonia methane	&	-0.418224310	&	-0.220748394	&	-0.196449504	\\
8	&	water - methane	&	-0.430386802	&	-0.233113226	&	-0.196415042	\\
9	&	formaldehyde dimer	&	-0.724913239	&	-0.361691753	&	-0.361440399	\\
10	&	water - ethene	&	-0.562480181	&	-0.327136401	&	-0.233408488	\\
11	&	formaldehyde - ethene	&	-0.689835019	&	-0.326859061	&	-0.361017399	\\
12	&	ethyne dimer	&	-0.588168890	&	-0.293741053	&	-0.293258671	\\
13	&	ammonia - ethene	&	-0.549360538	&	-0.326921792	&	-0.220684623	\\
14	&	ethene dimer	&	-0.656316561	&	-0.327002824	&	-0.326522937	\\
15	&	methane - ethene	&	-0.524290760	&	-0.326648269	&	-0.196429350	\\
16	&	borane - methane	&	-0.322172501	&	-0.196705744	&	-0.121985242	\\
17	&	methane - ethane	&	-0.560054059	&	-0.196642942	&	-0.361207343	\\
18	&	methane - ethane	&	-0.559365997	&	-0.196577997	&	-0.361162777	\\
19	&	methane dimer	&	-0.394435631	&	-0.196503395	&	-0.196503395	\\
20	&	Ar - methane	&	-0.366472187	&	-0.196315236	&	-0.169216520	\\
21	&	Ar - ethene	&	-0.496340364	&	-0.326243464	&	-0.169123725	\\
22	&	ethene - ethyne	&	-0.623414144	&	-0.326621275	&	-0.293173575	\\
23	&	ethene dimer	&	-0.657631984	&	-0.326689572	&	-0.326689572	\\
24	&	ethyne dimer	&	-0.589488331	&	-0.293125371	&	-0.293125371	\\
\hline
        \end{tabular}
        \end{table}

\subsection{Atomic orbital basis set: aug-cc-pVTZ}

\subsubsection{SVD amplitude threshold 10$^{-6}$}

\begin{table}[H]
	\centering
	\caption{Correlation energies resulting from SVD-DC-CCSDT+ calculations, using an MPFIT of aug-cc-pVTZ, a JKFIT of cc-pVQZ, an atomic orbital basis set of aug-cc-pVTZ and an SVD amplitude threshold of 10$^{-6}$ with counterpoise corrections. All values in Hartree.}
	\label{3}
	\begin{tabular}{clccc}
		\hline
	 ID &System &Dimer & Monomer 1 & Monomer 2 \\
		\hline
1	&	water - ammonia	&	-0.547047212	&	-0.283067000	&	-0.261212478	\\
2 &	water dimer	&	-0.567392122	&	-0.282825274	&	-0.282745977	\\
3 &	HCN dimer	&	-0.748315599	&	-0.373525550	&	-0.373983907	\\
4 &	HF dimer	&	-0.579938474	&	-0.289517735	&	-0.289544281	\\
5 &	ammonia dimer	&	-0.524738170	&	-0.261125933	&	-0.261125933	\\
6 &	HF - methane	&	-0.519769841	&	-0.228250041	&	-0.289450928	\\
7 &	ammonia - methane	&	-0.490427151	&	-0.261060666	&	-0.228160103	\\
8 &	water - methane	&	-0.511645678	&	-0.282485113	&	-0.228148665	\\
9 &	formaldehyde dimer	&	-0.864862178	&	-0.431103889	&	-0.431001794	\\
10 &	water - ethene	&	-0.666051231	&	-0.380837850	&	-0.282730520	\\
11&	formaldehyde - ethene	&	-0.813768318	&	-0.380682762	&	-0.430713203	\\
12	&	ethyne dimer	&	-0.689016440	&	-0.343762623	&	-0.343770283	\\
13	&	ammonia - ethene	&	-0.643855013	&	-0.380699816	&	-0.261101611	\\
14&	ethene dimer	&	-0.764342119	&	-0.380708175	&	-0.380567560	\\
15	&	methane - ethene	&	-0.610045064	&	-0.380581443	&	-0.228142638	\\
16&	borane - methane	&	-0.371922373	&	-0.228241748	&	-0.139583361	\\
17	&	methane - ethane	&	-0.651726102	&	-0.228222192	&	-0.421082315	\\
18	&	methane - ethane	&	-0.651007854	&	-0.228181771	&	-0.421048787	\\
19&	methane dimer	&	-0.457913608	&	-0.228161455	&	-0.228161455	\\
20&	Ar - methane	&	-0.465799849	&	-0.228092703	&	-0.236597359	\\
21&	Ar - ethene	&	-0.617620358	&	-0.380270133	&	-0.236201200	\\
22	&	ethene - ethyne	&	-0.728080691	&	-0.380574298	&	-0.343600558	\\
23	&	ethene dimer	&	-0.765718499	&	-0.380595206	&	-0.380595206	\\
24&	ethyne dimer	&	-0.690708200	&	-0.343573481	&	-0.343573481	\\
\hline
	\end{tabular}
\end{table}

\subsubsection{SVD amplitude threshold 10$^{-7}$}

\begin{table}[H]
	\centering
	\caption{Correlation energies resulting from SVD-DC-CCSDT+ calculations, using an MPFIT of aug-cc-pVTZ, a JKFIT of cc-pVQZ, an atomic orbital basis set of aug-cc-pVTZ and an SVD amplitude threshold of 10$^{-7}$ with counterpoise corrections. All values in Hartree.}
	\label{4}
	\begin{tabular}{clccc}
		\hline
	ID & 	System  & Dimer & Monomer 1 & Monomer 2 \\
		\hline
1 &	water - ammonia	&	-0.546887923	&	-0.282976355	&	-0.261128052	\\
2 &	water dimer	&	-0.567218631	&	-0.282735321	&	-0.282655385	\\
3 &	HCN dimer	&	-0.748165761	&	-0.373445112	&	-0.373903523	\\
4 &	HF dimer	&	-0.579783322	&	-0.289428458	&	-0.289455582	\\
5 &	ammonia dimer	&	-0.524562616	&	-0.261042077	&	-0.261042077	\\
6 &	HF - methane	&	-0.519644347	&	-0.228190073	&	-0.289362552	\\
7 &	ammonia - methane	&	-0.490289088	&	-0.260977470	&	-0.228099480	\\
8 &	water - methane	&	-0.511500191	&	-0.282392074	&	-0.228087189	\\
9 &	formaldehyde dimer	&	-0.864623567	&	-0.430972370	&	-0.430870942	\\
10&	water - ethene	&	-0.665869487	&	-0.380731641	&	-0.282636454	\\
11&	formaldehyde - ethene	&	-0.813535191	&	-0.380577882	&	-0.430583798	\\
12&	ethyne dimer	&	-0.688860154	&	-0.343680311	&	-0.343691002	\\
13&	ammonia - ethene	&	-0.643671474	&	-0.380594683	&	-0.261018274	\\
14&	ethene dimer	&	-0.764140177	&	-0.380603002	&	-0.380463603	\\
15&	methane - ethene	&	-0.609886791	&	-0.380475579	&	-0.228081137	\\
16&	borane - methane	&	-0.371861609	&	-0.228187825	&	-0.139560088	\\
17&	methane - ethane	&	-0.651565671	&	-0.228163811	&	-0.420972180	\\
18&	methane - ethane	&	-0.650845619	&	-0.228121795	&	-0.420937666	\\
19&	methane dimer	&	-0.457793609	&	-0.228100744	&	-0.228100744	\\
20&	Ar - methane	&	-0.465725671	&	-0.228032508	&	-0.236578391	\\
21&	Ar - ethene	&	-0.617499329	&	-0.380165756	&	-0.236181865	\\
22&	ethene - ethyne	&	-0.727898888	&	-0.380469346	&	-0.343519531	\\
23&	ethene dimer	&	-0.765502218	&	-0.380489894	&	-0.380489894	\\
24&	ethyne dimer	&	-0.690519706 &	-0.343493696	&	-0.343493696	\\
\hline
	\end{tabular}
\end{table}